\pdfoutput=1

\documentclass[%
 aip,
 amsmath,amssymb,
 reprint,%
]{revtex4-1}

\usepackage{graphicx}
\usepackage{dcolumn}
\usepackage{bm}

\usepackage[utf8]{inputenc}
\usepackage[T1]{fontenc}
\usepackage{mathptmx}
\usepackage{mathrsfs}

\begin{document}

\preprint{AIP/123-QED}

\title[Machine learning algorithms for predicting the amplitude of chaotic laser pulses]{Machine learning algorithms for predicting the amplitude of \\chaotic laser pulses}

\author{Pablo Amil}
\email{pamil@fisica.edu.uy}
 \affiliation{Departament de Física, Universitat Politécnica de Catalunya, St. Nebridi 22, Terrassa 08222, Barcelona, Spain.}

\author{Miguel C. Soriano}%
\affiliation{ 
Instituto de F\'isica Interdisciplinar y Sistemas Complejos, IFISC (CSIC-UIB), Campus Universitat de les Illes Balears, E-07122 Palma de Mallorca, Spain.
}%

\author{Cristina Masoller$^1$}

\date{\today}

\begin{abstract}
Forecasting the dynamics of chaotic systems from the analysis of their output signals is a challenging problem with applications in most fields of modern science. In this work, we use a laser model to compare the performance of several machine learning algorithms for forecasting the amplitude of upcoming emitted chaotic pulses. We simulate the dynamics of an optically injected semiconductor laser that presents a rich variety of dynamical regimes when changing the parameters. We focus on a particular dynamical regime that can show ultra-high intensity pulses, reminiscent of rogue waves. We compare the goodness of the forecast for several popular methods in machine learning, namely deep learning, support vector machine, nearest neighbors and reservoir computing. Finally, we analyze how their performance for predicting the height of the next optical pulse depends on the amount of noise and the length of the time-series used for training.
\end{abstract}

\maketitle

\begin{quotation}
Predicting the dynamical evolution of chaotic systems is an extremely challenging problem with important practical applications. With unprecedented advances in computer science and artificial intelligence, many algorithms are nowadays available for time series forecasting. Here, we use a well-known chaotic system of an optically injected semiconductor laser that exhibits fast and irregular pulsing dynamics to compare the performance of several algorithms (deep learning, support vector machine, nearest neighbors and reservoir computing) for predicting the amplitude of the next pulse. We compare the predictive power of such machine learning methods in terms of data requirements and the robustness towards the presence of noise in the evolution of the system. Our results indicate that an accurate prediction of the amplitude of upcoming chaotic pulses is possible using machine learning techniques, although the presence of extreme events in the time series and the consideration of stochastic contributions in the laser model bound the accuracy that can be achieved.
\end{quotation}

\section{\label{sec:Intro}Introduction}
Optically injected semiconductor lasers have a rich variety of dynamical regimes, including stable locked emission, regular pulsing and chaotic behavior \cite{ohtsubo2012semiconductor,wieczorek2005dynamical}. These regimes have found several practical applications. For example, under stable emission the laser emits light at the injected wavelength (the so-called injection-locking region) and has a high resonance frequency and a large modulation bandwidth \cite{lau2008strong}, which have broad applications for optical communications. The regular pulsing regime can be used for microwave generation \cite{lo2017numerical,xue2018narrow}, while the broad-band chaotic signal can be exploited for ultra-fast random number generation \cite{li2013heterodyne}. 

In turn, the output of the laser in the chaotic regime can be used for testing new methods for data analysis, and in particular, for time series prediction. Predicting the dynamical evolution of complex systems from the analysis of their output signals is an important problem in nonlinear science \cite{kollisch2018nonlinear,franzke2012predictability,birkholz2015predictability}, with a wide range of interdisciplinary applications. In these ``big data'' days, a significant number of researchers are focusing on developing novel methods for time series forecasting based on machine learning algorithms \cite{pathak2018model,isensee2019predicting,bialonski2015data,kuremoto2014time,wang2011chaotic}.

Delay embedding and recurrent neural networks have been used to predict the evolution of chaotic systems such as the Lorenz system and the Mackey-Glass system \cite{ardalani2010chaotic}. Locally linear neurofuzzy models~\cite{gholipour2006predicting} and support vector machine~\cite{lau2008local} have also been used to forecast chaotic signals. Here, in contrast with previous works, we do not attempt to forecast the evolution of a chaotic system, but the amplitude of the next peak in the observed signal.

As a case study, we consider the dynamics of an optically injected laser. We simulate the laser dynamics using a well-known rate equation model~\cite{ohtsubo2012semiconductor,perrone2014controlling}, and use the chaotic regime to compare the performance of several machine learning algorithms (deep learning, support vector machine, nearest neighbors and reservoir computing) for forecasting the amplitude of the next intensity pulse. 

Our main motivation to study this system is that it can be implemented experimentally and we hope that our work will motivate the analysis of real data. An important characteristic of this laser system is that it has control parameters (that can be varied in the experiment) that allow to generate time series with or without extreme pulses. 

Therefore, in the simulations, within the chaotic regime, we consider two different situations: the intensity pulses display occasional extreme values (so-called optical rogue waves \cite{akhmediev2016roadmap, bonatto2011deterministic}) or the intensity pulses are irregular but do not display extreme fluctuations. In the first case, the probability distribution function (pdf) of pulse amplitudes is long tailed, while in the second case, it has a well-defined cut off. 

The possibility of predicting and suppressing extreme pulses in a chaotic system has been demonstrated in \cite{cavalcante2013predictability}, but in this work the authors did not attempt to predict the pulse amplitude but rather the occurrence of a very high pulse whenever the trajectory approached a particular region of the phase space. To shed light on the limits of the forecast of extreme events, we consider dynamical regimes with and without extreme pulses, produced by the same underlying system, and we attempt to predict the amplitude of the next pulse, regardless of whether it is normal or extreme. In our system we find that, while both regular and extreme pulses can be forecasted, the existence of extreme pulses bounds the prediction accuracy. In an experimental setup, observational noise and the limited bandwidth of the detection system (photodiode, oscilloscope) can further limit the predictability of the pulse amplitude.

\section{Model} \label{sec:Simulations}

We simulated the dynamics of the complex optical field $E$ and the carrier population $N$ in a semiconductor laser with optical injection using the following rate equations \cite{wieczorek2005dynamical,zamora2013rogue}.

\begin{eqnarray}
\frac{dE}{dt}=\kappa\left(1+i\text{\ensuremath{\alpha}}\right)\left(N-1\right)E+\nonumber\\
+i\Delta\omega E+\sqrt{P_{inj}}+\sqrt{D}\xi\left(t\right)\;\label{eq:one},
\\
\frac{dN}{dt}=\gamma_{N}\left(\mu-N-N\left|E\right|^{2}\right)
\label{eq:two}.
\end{eqnarray}

The parameters in Eqs.~\ref{eq:one}-\ref{eq:two}  are: $\kappa$, the field decay rate, which we set at 300 ns$^{-1}$; $\alpha$, the linewidth enhancement factor, which we set at 3; $\Delta\omega$, the optical frequency detuning, which we set at $2\pi\times0.49$ GHz; $P_{inj}$, the optical injection strength, which we set to 60 ns$^{-2}$; $D$, the noise level, which we varied; $\gamma_N$, the carrier decay rate, which we set at 1 ns$^{-1}$, $\mu$ the pump current parameter, which we varied. $\xi(t)$ is a complex uncorrelated Gaussian noise of zero mean and unity variance that represents spontaneous emission: $\xi(t)=\xi_r(t)+i\xi_i(t)$ with $\left<\xi_r(t)\xi_r(t')\right>=\delta(t-t')$, $\left<\xi_i(t)\xi_i(t')\right>=\delta(t-t')$ and $\left<\xi_r(t)\xi_i(t')\right>=0$.

To simulate the evolution of Eqs. \ref{eq:one} and \ref{eq:two}, we used the Runge-Kutta method of order 2 with a time step of $10^{-3}$ ns, as described in \cite{san2000stochastic}, which takes into account the stochastic evolution with white noise.
In this work, we will analyze the chaotic pulses that appear at the output intensity of the laser defined as $P=|E|^2$.

\begin{figure} 
\includegraphics[width=\columnwidth]{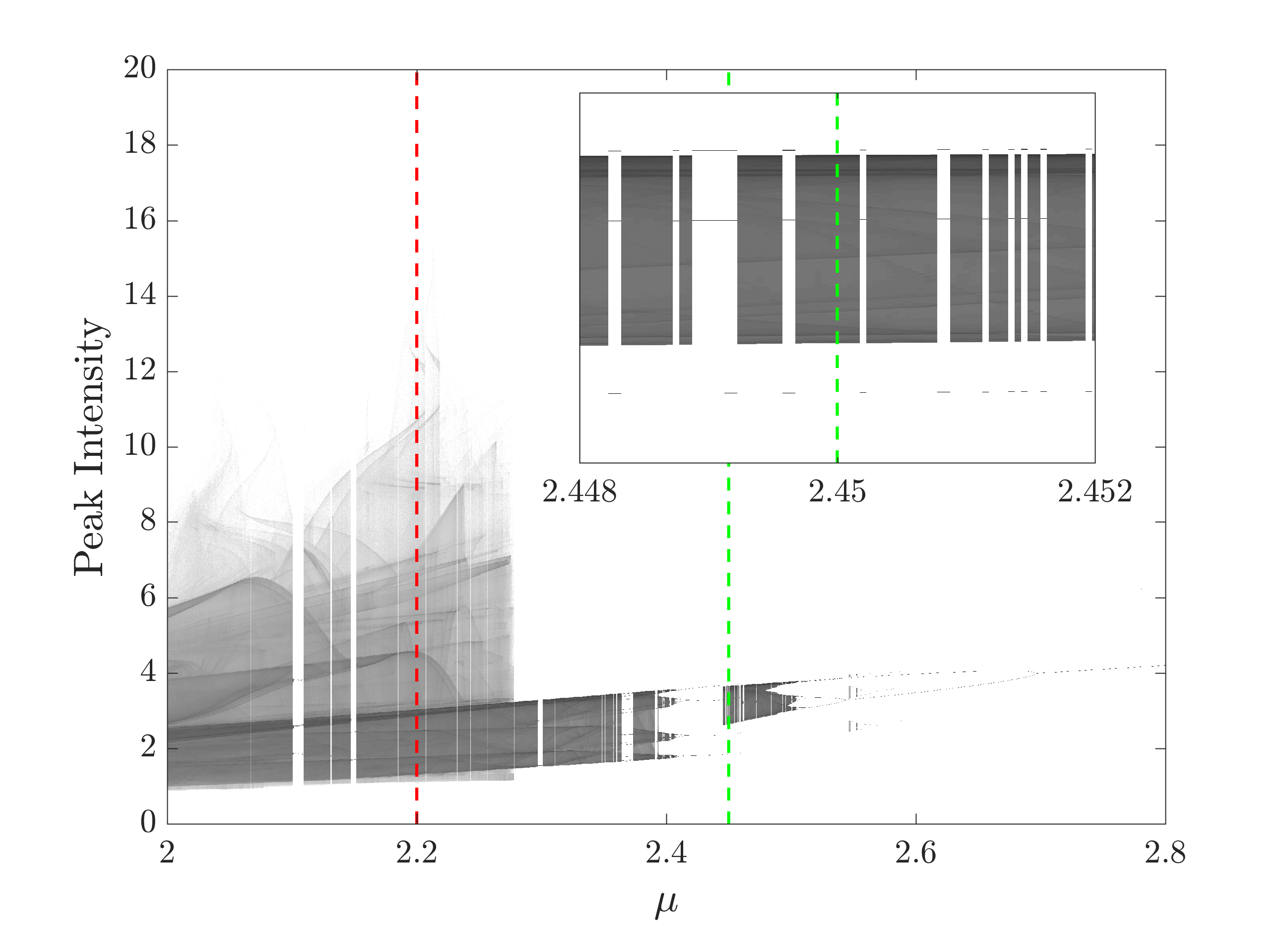}
\caption{\label{fig:MapaMU} Deterministic bifurcation diagram of the output intensity of the injected laser ($D$=0) when varying the pump current parameter $\mu$. For the subsequent analysis, we choose two currents that lead to very different dynamical behaviours: (i) $\mu=2.2$, where the system presents extreme events, and (ii) $\mu=2.45$ where the systems displays a bounded chaotic behaviour. Example time series for these parameters, including noise in the simulations ($D=10^{-4} \,\textrm{ns}^{-1}$) are shown in Fig. \ref{fig:EjemploSeries}.}
\end{figure}

\begin{figure} 
\includegraphics[width=0.95\columnwidth]{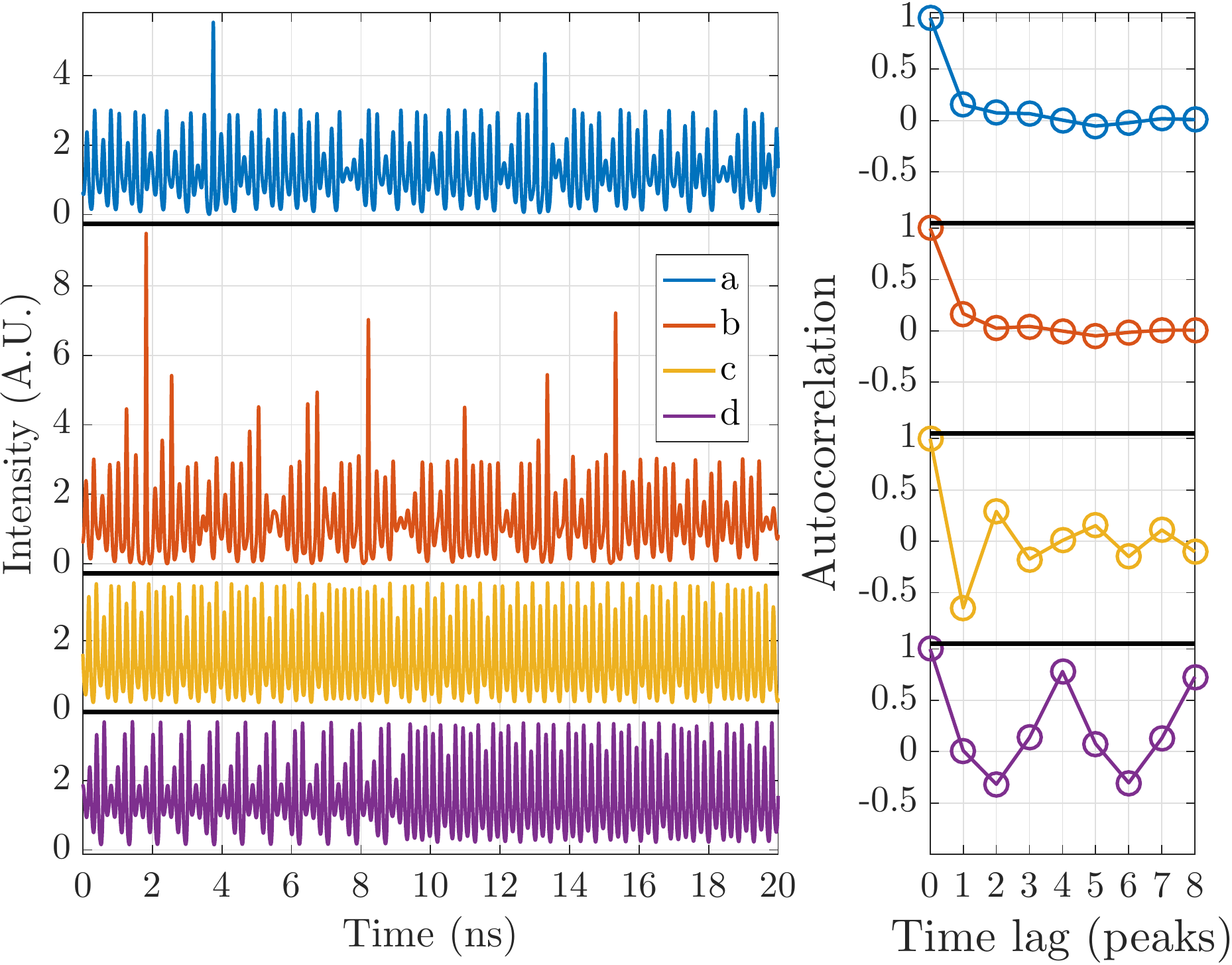}
\caption{\label{fig:EjemploSeries} (Left) Intensity time-series of the laser with optical injection and (right) autocorrelation function of the extracted peak series for $\mu=2.2$ (a and b) and $\mu=2.45$ (c and d) and noise level of $D=0$ (a and c) and $D=10^{-4} \,\textrm{ns}^{-1}$ (b and d).
}
\end{figure}

Figure \ref{fig:MapaMU} displays how the intensity deterministic dynamics ($D=0$) depends on the pump current parameter $\mu$. For small $\mu$ (not shown), the laser emits a constant intensity, but as $\mu$ increases a Hopf bifurcation and a series of period-doubling bifurcations occur, resulting in chaotic emission. Around $\mu=2.2$, the intensity shows extreme pulses as shown in Fig. \ref{fig:MapaMU} and the time series in Fig. \ref{fig:EjemploSeries}(a and b).
In contrast, at around $\mu=2.45$, the amplitude of the pulses in this chaotic regime is tightly bounded as it can be seen in Fig. \ref{fig:MapaMU} and in the time series in Fig. \ref{fig:EjemploSeries}(c and d).

The autocorrelation functions of the peak intensity values (i.e. the autocorrelation of the series $y_i$ built with the amplitude of each intensity peak) for both values of $\mu$ and both values of $D$ are shown in Fig. \ref{fig:EjemploSeries}. 
For $\mu=2.2$, the autocorrelation of the peak series decays to zero after a few peaks, both for $D=0$ and $D=10^{-4} \,\textrm{ns}^{-1}$.
It can be seen that for $\mu=2.45$, the autocorrelation of the peak series does not decay to zero and shows non-negligible values of the autocorrelation even after 8 peaks. The values of the autocorrelation as a function of the time lag (number of peaks) are larger for the series with noise.
We show in Fig. \ref{fig:EjemploSeries}(d) that the evolution of the laser intensity with noise alternates regions of more regular behaviour with regions of chaotic dynamics, which is not seen in the time series without noise in Fig. \ref{fig:EjemploSeries}(c). This is due to the fact that $\mu=2.45$ lies in a small chaotic island near regular regimes (see Fig. \ref{fig:MapaMU}) and we find noise-induced jumps between different dynamical regimes.
We anticipate that the faster decay of the autocorrelation function, together with the presence of extreme pulses, in the time series for $\mu=2.2$ will result on larger prediction errors than in the time series for $\mu=2.45$.

\section{Forecast methods} \label{sec:methods}

All machine learning methods used here tackle the problem of function approximation. We use them to forecast the amplitude of the upcoming intensity peaks by assuming that there is an objective function (that we try to infer) that takes as inputs a certain number of consecutive peak amplitudes and returns as output the amplitude of the next peak.

Except for the method of reservoir computing (that has an internal state with memory of the history of the inputs), all other methods are memoryless (i.e. they have no internal state of the history of the inputs), and explicit input and outputs of the objective function have to be provided in the training phase, providing information of the history with the previous intensity peaks amplitude. Let $y_i$ be the $i$-th intensity peak amplitude, our objective function is
\begin{equation}
f\left(y_{i-n},...,y_{i-1}\right)=y_{i}\label{eq:f1},
\end{equation}
\noindent where $n$ is the number of input intensity peak amplitudes that the machine learning algorithm is fed with. For the forecast of the peak amplitudes, we found that keeping $n=3$ yielded the minimum prediction error and further increasing $n$ produced no accuracy enhancement.
This choice will be justified in more detail in the results section (see Fig. \ref{fig:Nmaxs}).

For simplicity we call $\mathbf{x}_{i}=\left(y_{i-n},...,y_{i-1}\right)$ and thus, we can rewrite eq. \ref{eq:f1} as:
\begin{equation}
f\left(\mathbf{x}_{i}\right)=y_{i}\label{eq:f2}.
\end{equation}

For testing the methods, we use a different realization of the same simulations (not used in the training phase), and with this new data we evaluated the learned function,
\begin{equation}
\tilde{f}\left(\mathbf{x}_{i}\right)=\tilde{y}_{i}.
\end{equation}
Several statistical measures have been used in the literature to quantify the performance of time series prediction algorithms such as the correlation coefficient (CC) \cite{he2014comparative}, the mean squared error (MSE) \cite{ardalani2010chaotic}, the normalized mean squared error (NMSE) \cite{gholipour2006predicting}, the root mean squared error (RMSE) \cite{ardalani2010chaotic}, the normalized root-mean-square error (NRMSE) \cite{lau2008local}
, the mean absolute relative error (MARE), etc. Here we use the MARE \cite{he2014comparative} defined as: 
\begin{equation}
MARE=\frac{1}{N}\sum_{i=1}^{i=N}\frac{\left|\tilde{y}_{i}-y_{i}\right|}{y_{i}}\label{eq:MARE}.
\end{equation}


In the following subsections we describe the different algorithms used.

\subsection{Statistical methods}
\subsubsection{k-Nearest Neighbours}
The $k$-Nearest Neighbours (KNN) is a popular method used for supervised learning \cite{altman1992introduction}. It works by finding, in the training set, the k most similar points to a test point. Then, the prediction of the test point is obtained by averaging the response of such $k$ points (in the training set). Thus,
\begin{equation}
\tilde{y}=\frac{1}{k}\sum_{j\in\mathscr{N}}y_{j},
\end{equation}
where $\mathscr{N}$ (the neighbourhood of the test point  $\mathbf{x}_{i}$)  is the set of indexes of the $k$ points in the training set 
that are closest to the test point.

\subsubsection{Support Vector Machine}
Support Vector Machine \cite{boser1992training,vapnik2013nature,huang2006kernel} (SVM), is another popular method used for supervised learning, which is based on the inner product of points in the set to approximate the response function \cite{drucker1997support}. Nonlinearities can be introduced straight-forwardly by modifying the inner product function. For linear SVM the inner product of two points ($\mathbf{x}_{i}$ and $\mathbf{x}_{j}$) is calculated as

\begin{equation}
\left\langle \mathbf{x}_{i},\mathbf{x}_{j}\right\rangle =\mathbf{x}_{i}^{t}\mathbf{x}_{j},
\end{equation}

\noindent while nonlinearity can be introduced by using a Gaussian kernel to calculate the inner product, 

\begin{equation}
\left\langle \mathbf{x}_{i},\mathbf{x}_{j}\right\rangle =\exp\left(-\frac{\left\Vert \mathbf{x}_{i}-\mathbf{x}_{j}\right\Vert }{2\sigma^{2}}\right).
\end{equation}

The objective function, $\tilde{f}\left(\mathbf{x}_{i}\right)=\tilde{y}_{i}$, is written as a linear combination of the inner products with the support vectors

\begin{equation}
    \tilde{f}\left(\mathbf{x}_{i}\right)=\sum_{j}\beta_{j}\left\langle \mathbf{x}_{j},\mathbf{x}_{i}\right\rangle +b.
\end{equation}

\noindent The coefficients $\beta_j$ and $b$ are obtained by solving a convex optimization problem \cite{vapnik2013nature}. 

The linear SVM has the advantage of being parameter-free. In contrast, for using the Gaussian kernel the scale factor $\sigma$ has to be defined. To set the value of $\sigma$ we used the automatic heuristic implemented in the {\it Statistics and Machine Learning Toolbox} of {\it Matlab} (the {\it fitrsvm} function).


\subsection{Artificial neural networks}

\subsubsection{Feed-forward neural networks}
Feed-forward neural networks, usually simply referred as neural networks, use a set of units, called perceptrons, that, when used in a large network, their output can approximate a great variety of functions depending on the weights of the connections among the units.

Perceptrons perform two tasks, they compute a weighted sum of all their inputs (and a constant bias input), and they perform a nonlinear function, called activation function, to the result. The output of the activation function is the output of the perceptron. Most commonly, the activation functions used are sigmoids, in this work we use the $\tanh$ function in all but the last (output) layer, in which we don't use a nonlinearity to avoid bounding the final output to the codomain of the nonlinearity.

A feed-forward neural network, is a network of such perceptrons wherein they are ordered in layers, as shown in Fig. \ref{fig:NN_RC}(a) for a single hidden layer. The perceptrons of the first layer have their inputs set to be the inputs of the whole network. For the rest of the layers, the inputs are defined as the outputs of the perceptrons in the previous layer.

The parameters of these networks are the weights of each perceptron. These parameters can be set using a gradient descend algorithm, in feed-forward neural networks an efficient algorithm to perform gradient descend, called back-propagation \cite{lecun1990handwritten}, may be used.

We used a shallow neural network (shallow NN), consisting of a single hidden layer of 30 perceptrons and a deep neural network (deep NN) consisting of 5 hidden layers of 10, 20, 50, 25, and 10 perceptrons (ordered from the input layer to the output layer), respectively.


\subsubsection{Reservoir computing}

\begin{figure} 
\includegraphics[width=0.95\columnwidth]{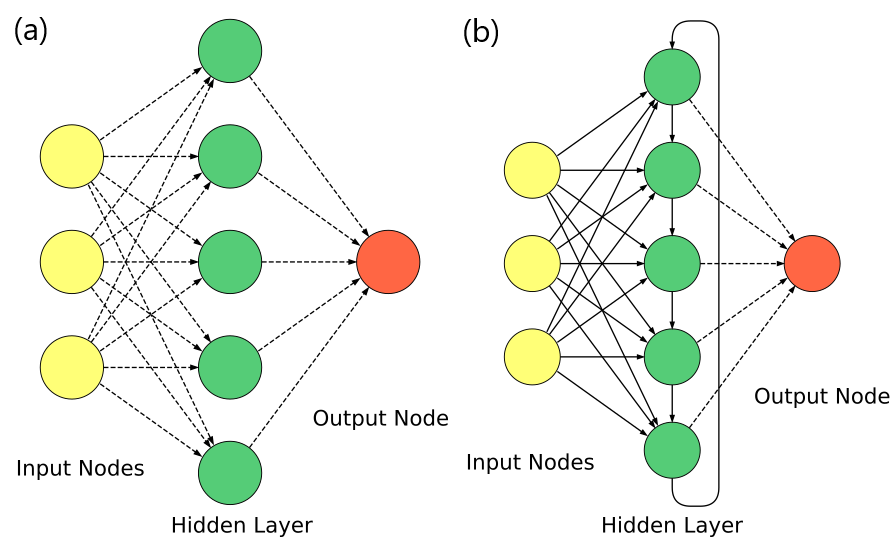}
\caption{\label{fig:NN_RC} Graphical representation of a feed-forward neural network with a single hidden layer (a) and a reservoir computer (b). The dashed lines represent the connections that can be adjusted via a learning procedure while the solid lines account for the connections that are randomly weighted and left untrained.}
\end{figure}

Reservoir computing (RC) is a computational paradigm that can be viewed as a particular type of artificial neural networks with a single hidden layer and recurrent connections \cite{verstraeten2007experimental}.
A ring topology in the hidden layer (or reservoir), as the one shown in Fig. \ref{fig:NN_RC}(b), is a simple way to create recurrent connections.
Such a ring topology yields a performance comparable to more complex network topologies in the reservoir \cite{rodan2010minimum}.
Being a recurrent neural network, the reservoir computing technique is suitable to process sequential information.
In reservoir computing, the connection weights from the input layer to the hidden layer as well as the connection weights within the reservoir are drawn from a Gaussian distribution and left untrained.
The connection weights from the reservoir to the output layer are trained in a supervised learning procedure, which translates to a linear problem that can be solved via a simple linear regression \cite{lukovsevivcius2009reservoir}.

The nodes in the reservoir layer perform a nonlinear transformation of the input data. 
Here we use a sine squared nonlinearity, which can be implemented in photonic hardware \cite{larger2012photonic,paquot2012optoelectronic}, but other types of nonlinearity are also possible.
Finally, the output node performs a weighted sum of the reservoir outputs.

The RC method can be described by the following equations for the states of the nodes in the hidden layer ($z^j$) and the prediction of the output node ($\tilde{y}$):

\begin{eqnarray}
z_i^j=F(\gamma w_j^I y_{i-1} + \beta z_{i-1}^{j-1}),\label{eq:RC1}
\\
\tilde{y}_i=\sum_{j=1}^{D}w_j^O z_i^j;
\label{eq:RC2}
\end{eqnarray}
where $i$ refers to the peaks in the laser time series, $j$ is the index of the node in the hidden layer, $w^I$ are the set of input weights drawn from a random Gaussian distribution, $\gamma$ and $\beta$ are the input and feedback scaling, respectively, and $F(u)=sin^2(u+\phi)$ is the nonlinear activation function. In eq. \ref{eq:RC2}, $D$ is the number of hidden nodes and $w^O$ stands for the trained output weights. In order to create a recurrent ring connectivity in the hidden layer (also known as reservoir), we connect node $z^j$ ($j=\{2...D\}$) with its neighbour $z^{j-1}$ and we close the ring by connecting node $z^1$ with $z^D$ as  shown in Fig. \ref{fig:NN_RC}(b). Here, we have set the hyper-parameter values as $\gamma=4.5$, $\beta=0.25$, $\phi=0.6\pi$ and $D=6000$, which minimize the prediction error. We have verified that a $tanh$ activation function yields quantitatively similar results once the hyper-parameters $\gamma$ and $\beta$ are optimized.

We note that the heuristic for the RC practitioners is to assume a random interconnection topology in the reservoir, which usually yields good results. However, regular network topologies also yield optimal results as long as the hyper-parameters are optimized \cite{kawai2019small,griffith2019forecasting}, as it has been the case here. 
For the RC method, we only feed a single amplitude value to predict the amplitude of the next pulse. Feeding the RC method with the value of several previous peaks would mean that, in practice, the reservoir computer would not need to use its own internal memory. The motivation to employ a different number of input peaks for the reservoir computer lies on the observation that it can reach a prediction error comparable to the other methods without using explicit memory of the preceding peaks.

\section{Results and discussion} \label{sec:Results}
We now proceed to evaluate the performance of the different forecast methods on the prediction of the amplitude of chaotic laser pulses.
The goal of our work is to predict the amplitude of the upcoming laser pulse given the recent history of the dynamics.
To that end, we generate long time series of a laser subject to optical injection following the model described in Eqs. (\ref{eq:one}) and (\ref{eq:two}) for the two chaotic regimes shown in Fig. \ref{fig:EjemploSeries}.

\begin{figure}[!h] 
\includegraphics[width=0.8\columnwidth]{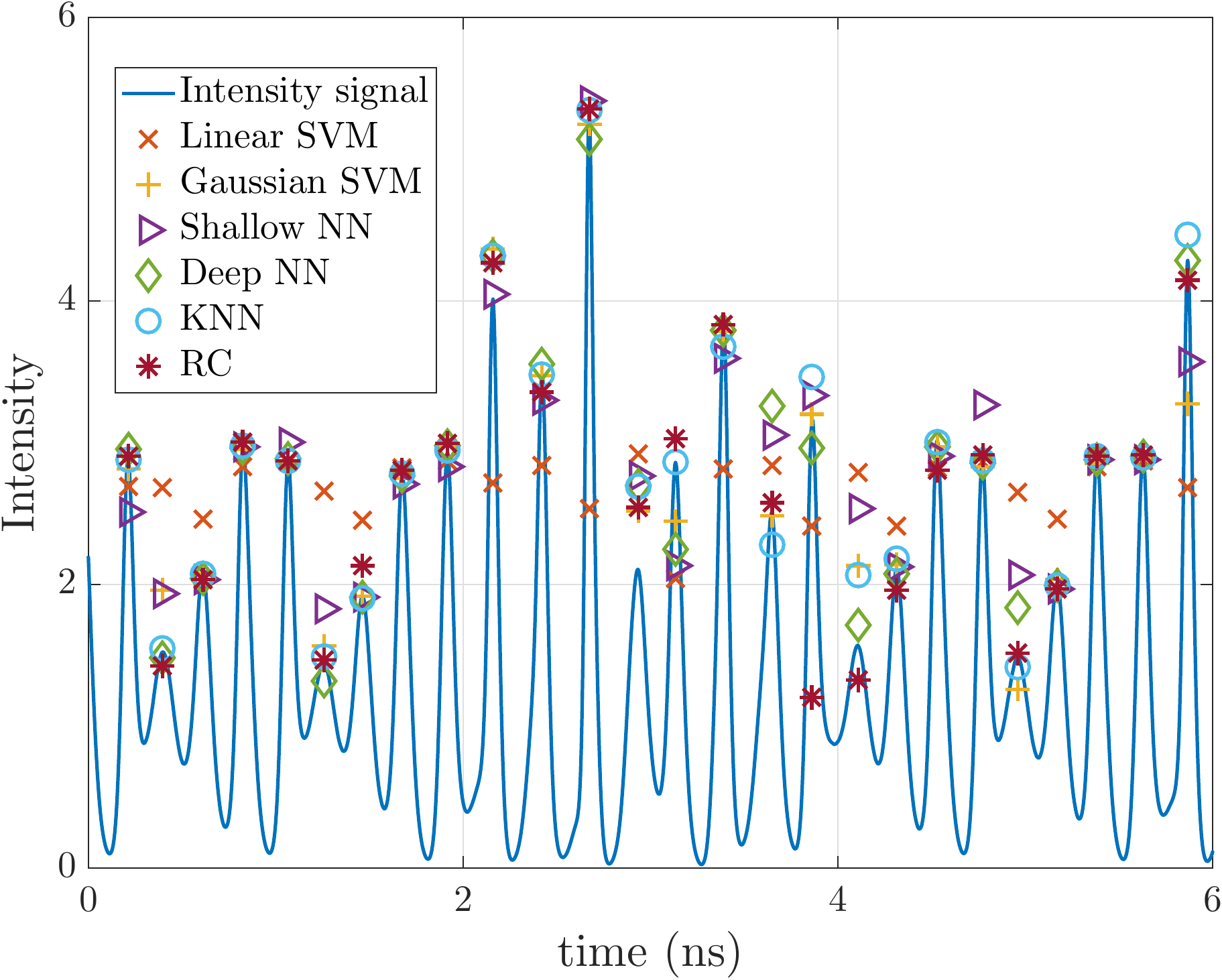}
\caption{\label{fig:EjemploPrediccion} Simulated intensity time series together with the peak amplitudes predicted by the different methods. The parameters are $D=10^{-4} \,\textrm{ns}^{-1}$ and $\mu=2.2$. All methods were trained using 15000 ns of simulation, which contain 65534 peak intensity values.}
\end{figure}

\begin{figure}[!h] 
\includegraphics[width=\columnwidth]{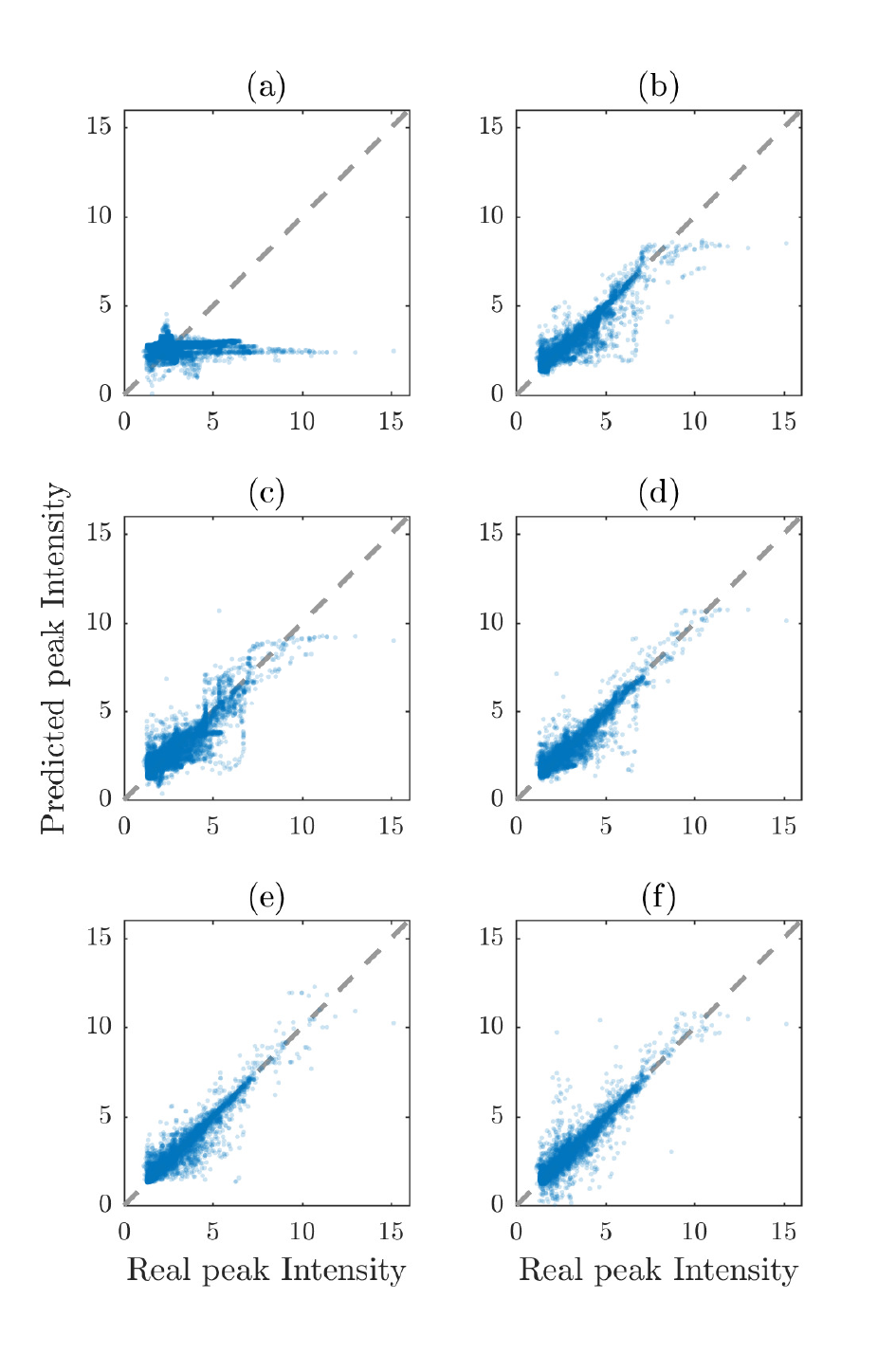}
\caption{\label{fig:Scatter} Scatter plots displaying the simulated peak intensity vs. the predicted peak intensity for the methods (a) Linear SVM, (b) Gaussian SVM, (c) Shallow Neural Network, (d) Deep Neural Network, (e) k-Nearest Neighbors, (f) Reservoir Computing. The parameters are $D=10^{-4} \,\textrm{ns}^{-1}$ and $\mu=2.2$. All methods were trained using 15000 ns of simulation, containing 65534 peak intensity values.
}
\end{figure}

By looking at Fig. \ref{fig:EjemploSeries}, the presence of extreme events in the time series of the laser when the current is $\mu=2.2$ becomes apparent.
We anticipate that the existence of such extreme events poses a challenge for the prediction of the chaotic laser pulses' amplitude. 
Fig. \ref{fig:EjemploPrediccion} shows a segment of the time series of the laser for the parameters $\mu=2.2$ and $D=10^{-4} \,\textrm{ns}^{-1}$ together with the prediction of the pulses amplitude for all the methods considered in this work.
From this first qualitative evaluation of the forecast methods, we can observe how the linear SVM method is outperformed by the other methods.
In turn, the methods Deep NN, KNN and RC tend to yield a similar, accurate, prediction of the amplitude of the chaotic pulses.

A further visualization of the goodness of the different methods is provided by the scatter plots in Fig.  \ref{fig:Scatter}.
These scatter plots represent the predicted peak intensities versus the real ones.
The methods with a better prediction accuracy need to align to a diagonal line in this representation.
For this chaotic regime of the laser dynamics with the presence of extreme events, the Deep NN, KNN and RC methods are well aligned to the diagonal lines as shown in Figs. \ref{fig:Scatter}(d)-(f).
In contrast, the Shallow NN and Gaussian SVM methods tend to underestimate the amplitude of medium to large pulses as it can be seen in \ref{fig:Scatter}(b)-(c).
As shown in Fig. \ref{fig:Scatter}(a), the linear SVM method fails to capture the complexity of the dynamics.

\begin{figure}[!h]
\begin{center}
\begin{tabular}{c}
\includegraphics[width=\columnwidth]{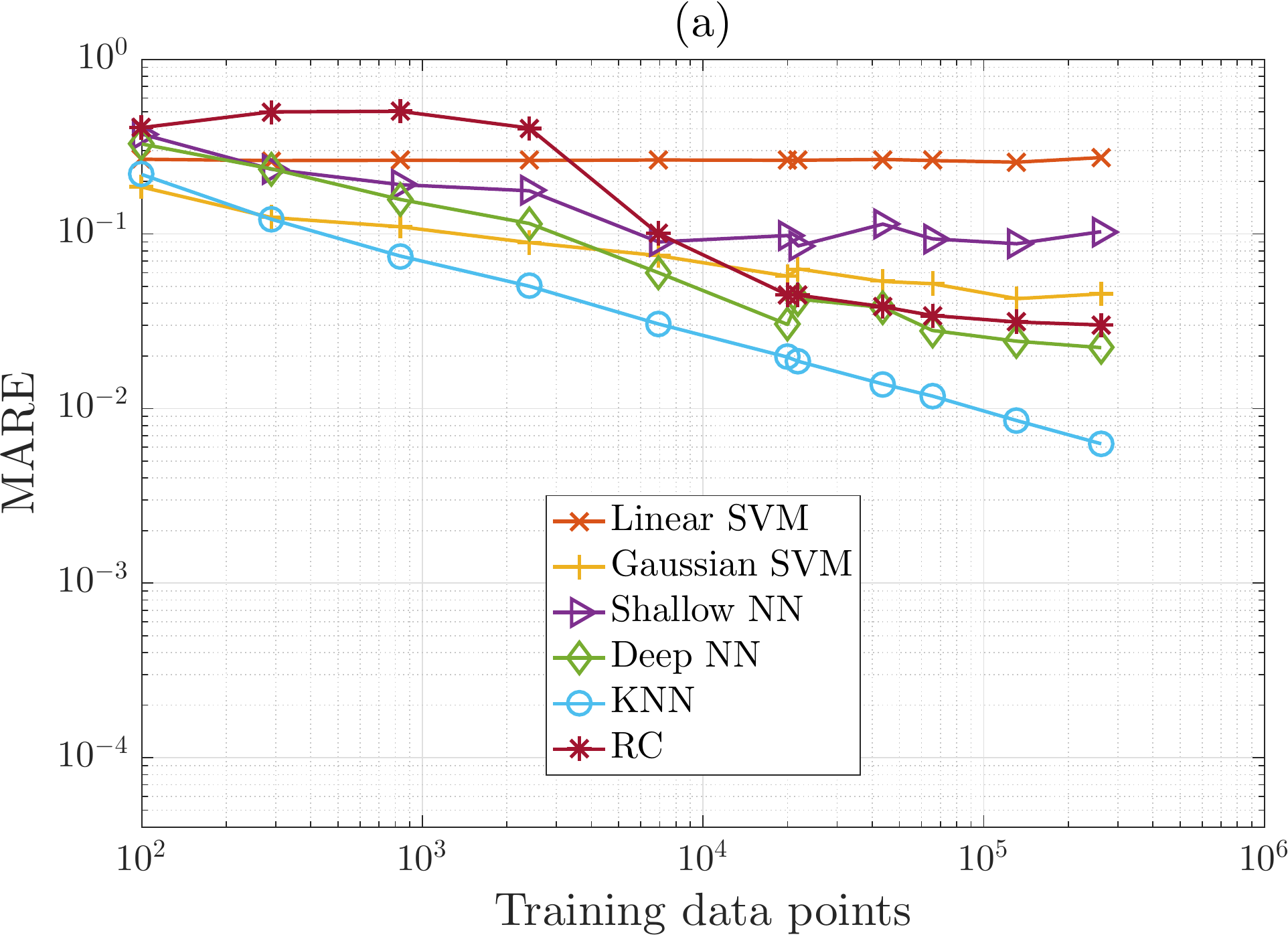} \tabularnewline
\includegraphics[width=\columnwidth]{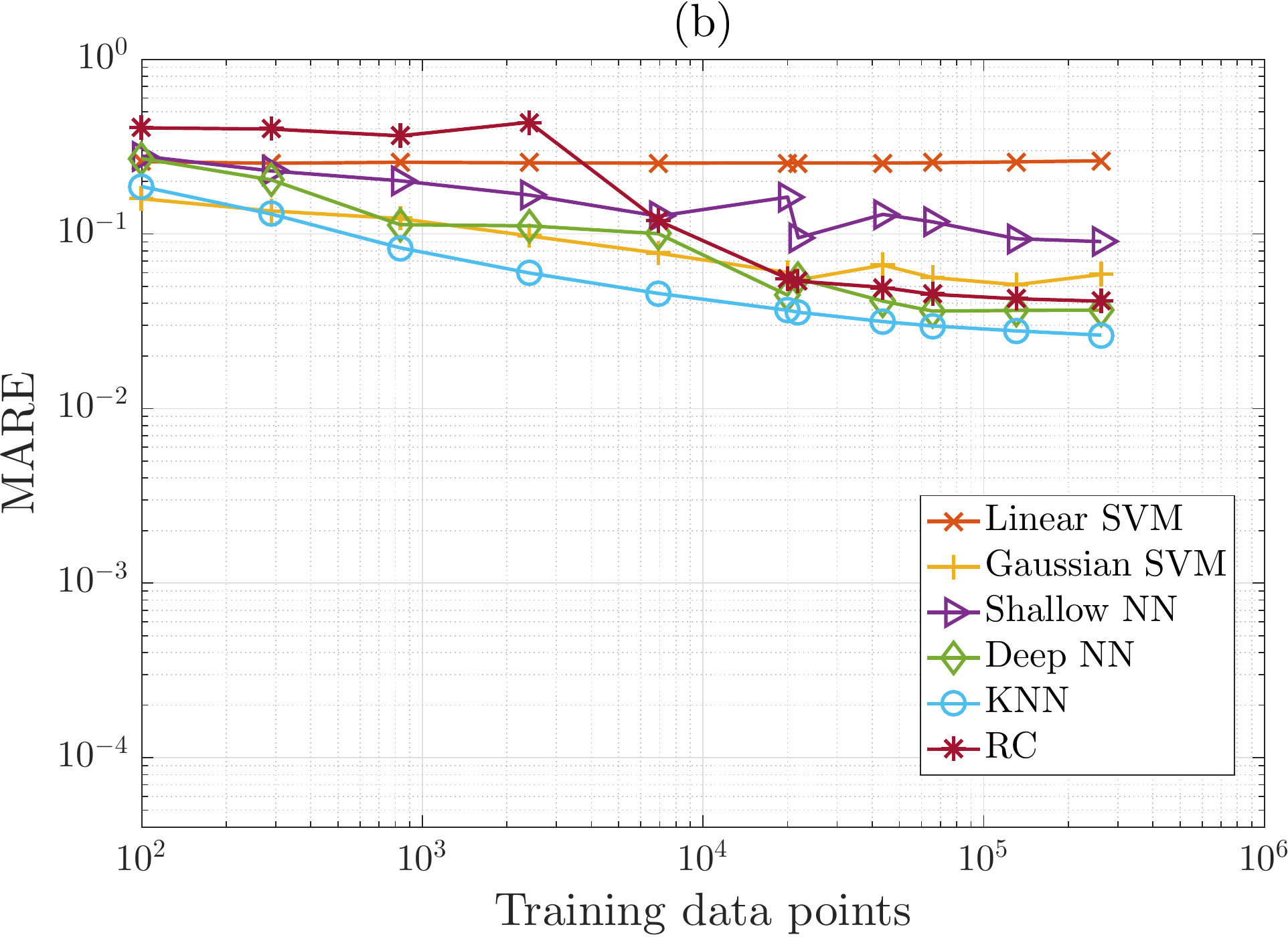} \tabularnewline
\end{tabular}
\end{center}
\caption{\label{fig:ErrorNpoints_2_2} Mean absolute relative error as a function of number of training points. We show the error of the peak amplitud prediction as a function of the number of training points for noise levels of (a) $D=0$ and (b) $D=10^{-4} \,\textrm{ns}^{-1}$, at $\mu=2.2$.}
\end{figure}

\begin{figure}[!h]
\begin{center}
\begin{tabular}{c}
\includegraphics[width=\columnwidth]{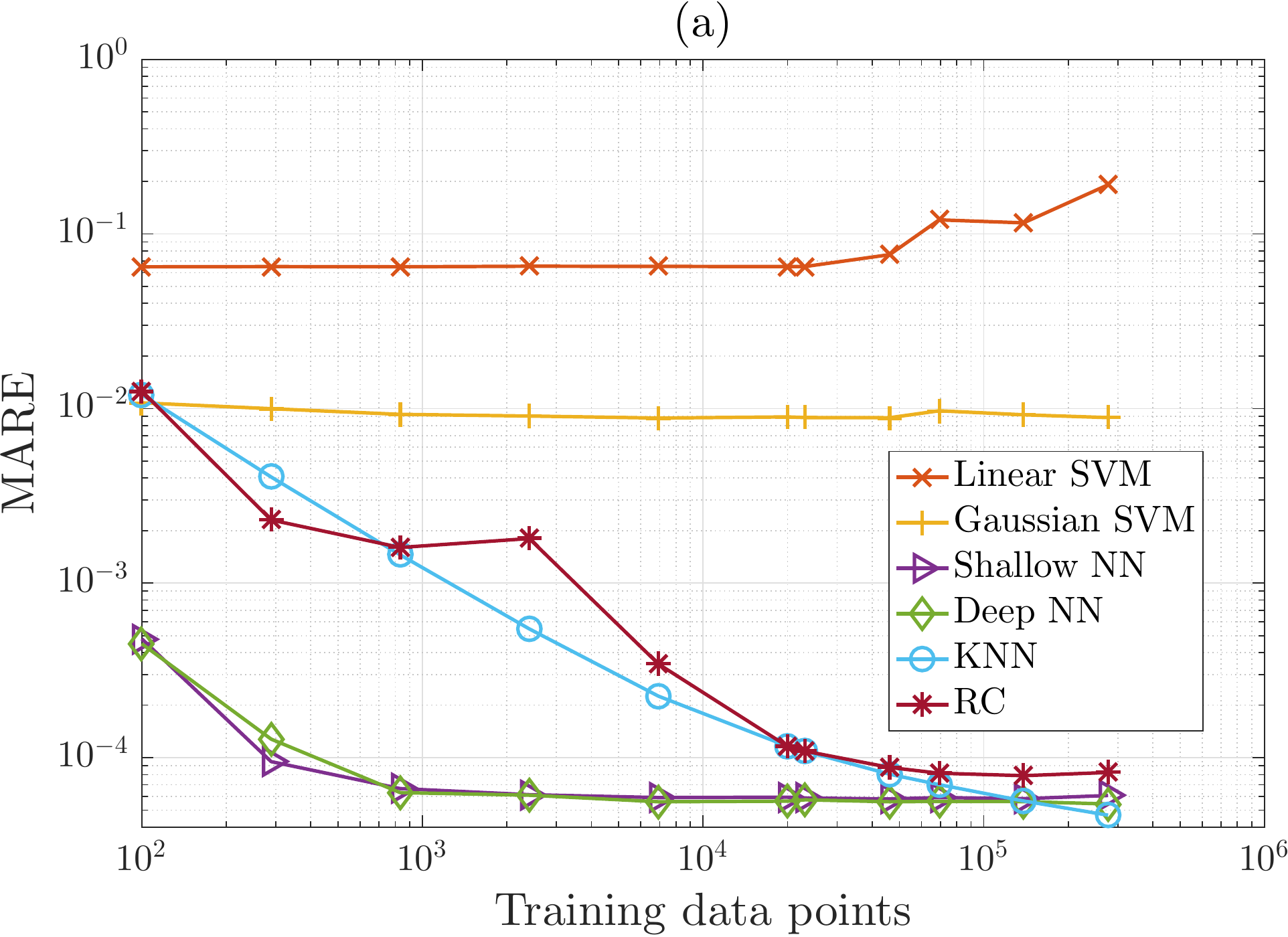} \tabularnewline
\includegraphics[width=\columnwidth]{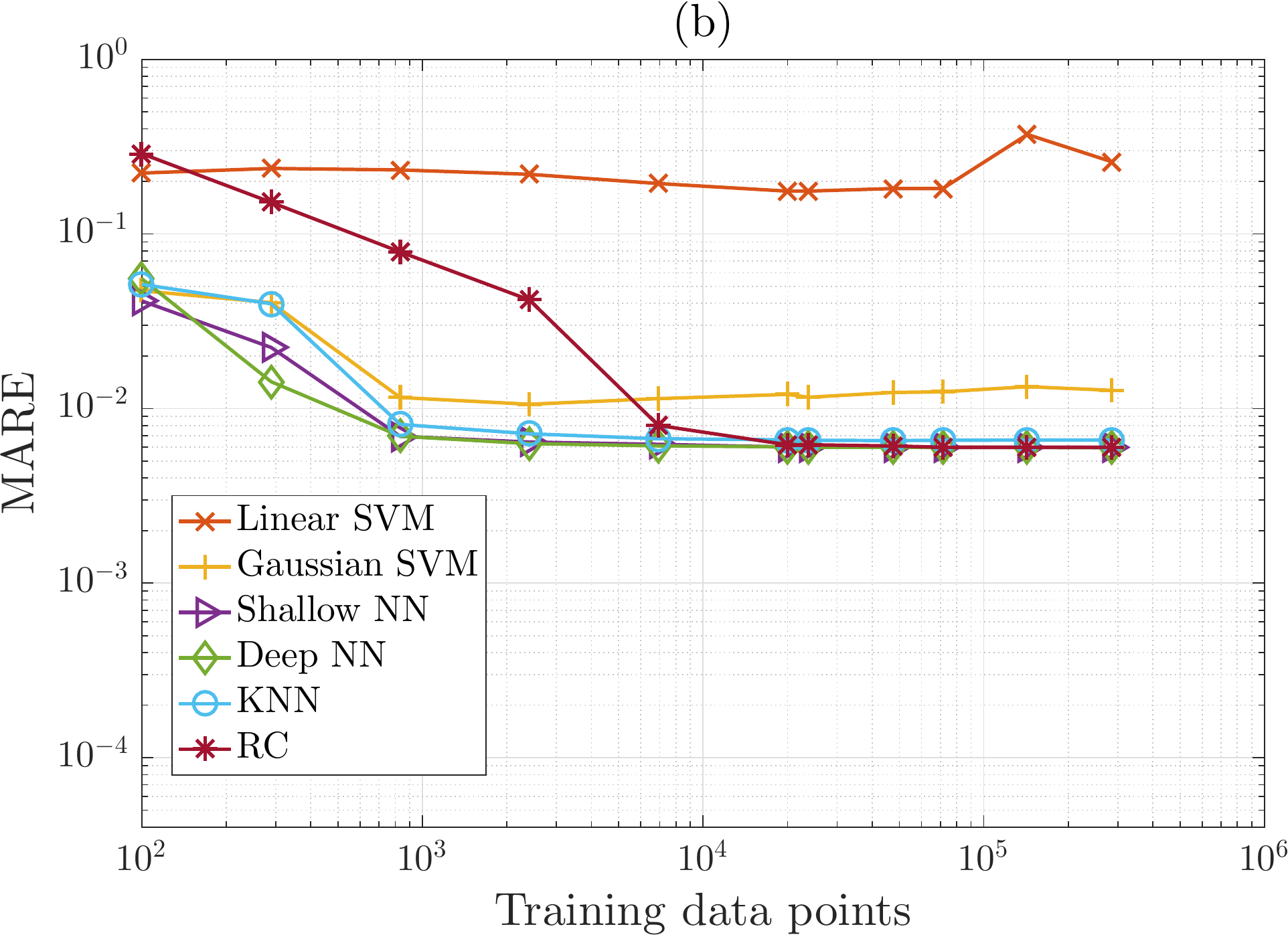} \tabularnewline
\end{tabular}
\end{center}
\caption{\label{fig:ErrorNpoints_2_45} Mean absolute relative error as a function of number of training points. We show the error of the peak amplitud prediction as a function of the number of training points for noise levels of (a) $D=0$ and (b) $D=10^{-4} \,\textrm{ns}^{-1}$, at $\mu=2.45$.}
\end{figure}

When doing data-driven forecasting, it is necessary to evaluate the number of training points needed to have accurate results. In Figs. \ref{fig:ErrorNpoints_2_2} and \ref{fig:ErrorNpoints_2_45} we show how the accuracy (as measured by the mean absolute relative error, Eq. \ref{eq:MARE}) depends on the number of points used to train the algorithms, when there are extreme pulses (Fig. \ref{fig:ErrorNpoints_2_2}) and when there are no extreme pulses (Fig. \ref{fig:ErrorNpoints_2_45}). 

First, we compare the forecast results for the noise-free numerical simulations at currents $\mu=2.2$ and $\mu=2.45$, which are shown in Figs. \ref{fig:ErrorNpoints_2_2}(a) and \ref{fig:ErrorNpoints_2_45}(a).
The MARE of the forecast for $\mu=2.2$ is at least two orders of magnitude worse than the forecast for $\mu=2.45$.
This is due to the added complexity of the extreme events at $\mu=2.2$, deteriorating the performance of all the forecasting methods.
We find that the KNN, Deep NN, and RC methods, in this order, yield the most accurate predictions for $\mu=2.2$.
These methods, together with the Shallow NN, yield the lowest MARE for $\mu=2.45$.
In both cases, the performance of the RC method becomes more accurate when the number of training data points is larger than the number of nodes in the reservoir ($D=6000$).
Overall, the prediction of the amplitude of the upcoming chaotic pulse for $\mu=2.45$ requires less training points than for $\mu=2.2$.
These results suggest that the forecast of the dynamics with extreme pulses is intrinsically harder to predict.
It could also be that the low frequency of the extreme pulses makes them more difficult to predict because they appear less frequently in the training set.
However, they also appear less frequently in the testing set and thus have less weight in the overall error.

Second, we analyze the influence of the stochastic contribution in Eq. \ref{eq:one} on the forecast of the pulses' amplitude.
We show in Figs. \ref{fig:ErrorNpoints_2_2}(b) and \ref{fig:ErrorNpoints_2_45}(b) that the presence of noise triggers an early plateau that bounds the MARE, deteriorating the performance of all the methods.
The stochastic contribution to the dynamics has a stronger influence on the forecast for the chaotic dynamics generated at $\mu=2.45$, with an increase of two orders of magnitude in the MARE as shown in Figs. \ref{fig:ErrorNpoints_2_45} (a) and (b).
When noisy dynamics is considered, the MARE for $\mu=2.45$ and $\mu=2.2$ are less than an order of magnitude apart (see Figs. 6 (b) and 7(b)) in contrast to the noise-free counterparts for which the difference in MARE between $\mu=2.45$ and $\mu=2.2$ is more apparent (see Figs. 6 (a) and 7(a)). The deterioration of the prediction accuracy in the presence of observational noise has also been reported e.g. in \cite{gholipour2006predicting}, where the NMSE decreased 5 to 6 orders of magnitude. 


\begin{figure} 
\includegraphics[width=\columnwidth]{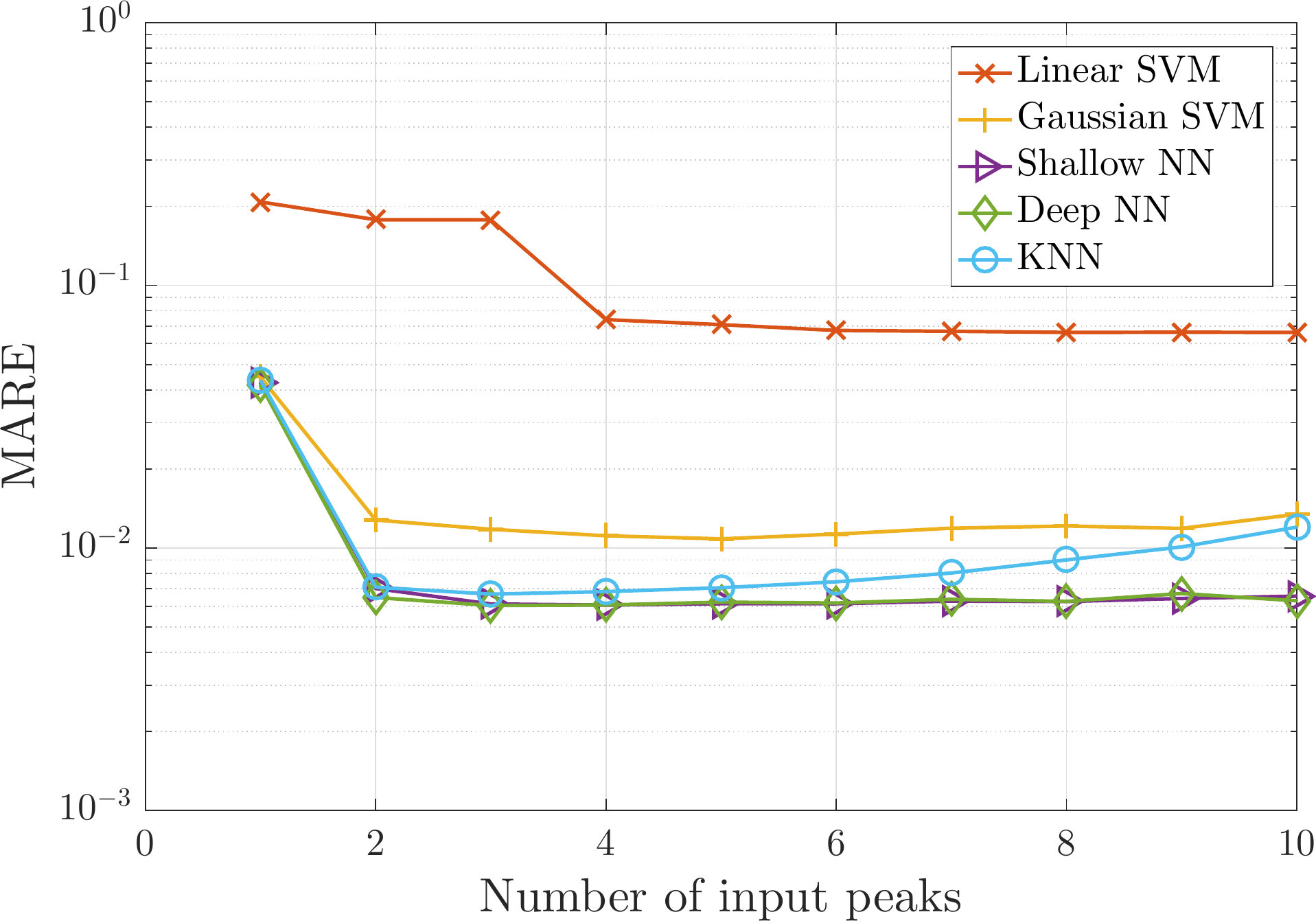}
\caption{\label{fig:Nmaxs} Mean absolute relative error as a function of the number of preceding pulses fed as the input to each algorithm. For this example, all algorithms were trained using 10000 data points at $\mu=2.45$ and $D=10^{-4} \,\textrm{ns}^{-1}$. For both KNN and Deep NN the minimum error occurs when using 3 input peaks, being the latter the absolute minimum in this plot considering all other methods.}
\end{figure}

An important parameter for all but the reservoir computing approach is the number of input intensity peak amplitudes ($n$) wherewith the machine learning algorithm is fed. 
This parameter sets the amount of history that the algorithm is able to ``see''.
In Fig. \ref{fig:Nmaxs}, we show how the performance changes when changing $n$ in the case of the chaotic dynamics with $\mu=2.45$ and $D=10^{-4} \,\textrm{ns}^{-1}$.
We used 10000 training data points to be well inside the plateau of performance seen in Fig. \ref{fig:ErrorNpoints_2_45}(b).
The results shown in this figure justifies our choice of using $n=3$, which yields a minimum MARE for most of the forecast methods.
For the RC method, we set $n=1$ as it is the only method that possesses an internal memory.

Another important issue to consider when implementing these data-driven methods is the computer power that is required to train and test each forecast method. 
Although different methods scale differently with the amount of data in the training set, some general rules of thumb apply.
In the KNN method, while there is no specific training time, the time for evaluating each test point, however, grows linearly with the amount of points in the dataset.
The KNN method is, in this sense, ideal for real-time data as it can take into account new data into the dataset without any extra computational overhead.
The computing power for training and testing the SVM methods depends greatly on the amount of support vectors that are needed, and on the kernel that is used. 
We find that the linear SVM method takes a greater time to train and a comparable time to test w.r.t. the Gaussian kernel method. 
This is due to the fact that the linear SVM method fails to capture the complexity of the data and, thus, a great amount of support vectors are needed.
The feed-forward neural networks and the RC method are (in respect to train and test) opposite to the KNN method, they take a great amount of time to train but are computationally cheap to evaluate test points.
The training time of the neural networks-based models depends on the length of the training data and on the amount of internal model parameters they have.
In our examples, the deep NN takes about 20 times the time of the shallow NN to train.
The RC method, on the other hand, has a simpler training mechanism, which takes approximately 5 times the time of the shallow NN to train.
All neural networks-based models take a comparable (low) time in the testing stage.

We end up with a comparison of the performances obtained here with the literature. The reported MARE values that have been obtained strongly vary with the algorithm used and the characteristics of the datasets analyzed. For example, machine learning techniques with delay embedding in real data give MARE values of the order of 0.15 for river flow prediction \cite{he2014comparative}, or as low as 0.025 for electricity consumption \cite{wang2011chaotic}. With time series simulated from the chaotic Ikeda map, a MARE value as low as $5.8\times10^{-5}$ was reported in  \cite{yang2009predict}. In general, the prediction of noisy (possibly chaotic) real-world dynamics yields larger errors than the prediction of synthetic numerical data without noise. A more precise direct comparison of previously published results is, however, not currently possible since we do not predict the future trajectory of the dynamics but the amplitude for the next pulse. 

\section{Conclusions} \label{sec:Conclusions}

We have used the chaotic dynamics of the intensity of an optically injected laser to test the performance of several machine learning algorithms for forecasting the amplitude of the next intensity pulse. This laser system is described by a simple model that, with a small change of parameters, produces time series which have extreme events in the form of high peak intensities, resembling the dynamics of much more complex systems. In spite of the fact that the autocorrelation function of the sequence of pulse amplitudes decays rapidly, good prediction accuracy was achieved with some of the proposed methods, namely the KNN, Deep NN and RC methods.
We have verified that the MARE for the most accurate methods (DNN, KNN and RC) remains approximately constant even for the prediction of extreme pulses that have a probability of appearance as low as 1/1000.

Our work suggests that similar methods may be used in the forecast of more complex systems, although further testing is of course necessary to asses how well they would perform in high-dimensional chaotic dynamical systems. While with simple dynamics, we only needed around 1000 data points to achieve maximum performance with some methods (shallow and deep NN); when forecasting more complex dynamics, some of the methods (KNN and deep NN) will continue improving their performance if longer datasets are available for training (longer than $10^5$ data points). 

We have also compared the performance of different variations of the same machine learning algorithm (compare linear to Gaussian SVM and shallow to deep neural networks in Figs. \ref{fig:EjemploPrediccion}-\ref{fig:ErrorNpoints_2_45}), especially relevant when considering big training datasets. We do not exclude that even more complex methods (e.g. a neural network with additional hidden layers) might outperform the 
presented algorithms.
However, the presented algorithms already serve the purpose of showing the dependence of the forecast error on the complexity of the dynamics and on the inclusion of stochastic contributions.

\begin{acknowledgments}
P. A. and C. M. acknowledge support by the BE-OPTICAL project (H2020-675512). C.M. also acknowledges support from Spanish Ministerio de Ciencia, Innovación y Universidades (PGC2018-099443-B-I00) and ICREA ACADEMIA. M.C.S. was supported through a ``Ramon y Cajal'' Fellowship (RYC-2015-18140). This work is the result of a collaboration established within  the Ibersinc network of excellence (FIS2017-90782-REDT).
\end{acknowledgments}


%

\end{document}